\documentclass[a4paper, 10pt]{article}

\usepackage{mathrsfs}
\usepackage{hyperref}
\usepackage{amsmath} 
\usepackage{epsfig,amsfonts} 
\usepackage[greek,english]{babel}
\usepackage{teubner}
\usepackage{graphicx}
\usepackage{orcidlink}
\newcommand{\be}{\begin{equation}}
\newcommand{\ee}{\end{equation}}
\newcommand{\bea}{\begin{eqnarray}}
\newcommand{\eea}{\end{eqnarray}}
\newcommand{\bem}{\begin{matrix}}
\newcommand{\eem}{\end{matrix}}
\newcommand{\nnb}{\nonumber}
\def\vac#1{{\bf #1}}
\def\koppa{\hbox{\foreignlanguage{greek}{\coppa}}}
\def\smallkoppa{{\hbox{\foreignlanguage{greek}{\footnotesize\coppa}}}}

\title{Ralph Kenna's scaling relations in critical phenomena}

\author{Le\"\i la Moueddene $^{1,2}$, Arnaldo Donoso $^{3}$ and Bertrand Berche  $^{1}$\thanks{Correspondence: bertrand.berche@univ-lorraine.fr}
  \orcidlink{0000-0002-4254-807X}}

\date{%
\small
   $^{1}$  Laboratoire de Physique et Chimie Th\'eoriques,  CNRS - Universit\'e de Lorraine, UMR 7019,
    	 Nancy, France 
	 and L4 collaboration, Leipzig-Lorraine-Lviv-Coventry, Europe\\
$^{2}$ {Centre for Fluid and Complex Systems, Coventry University, Coventry CV1 5FB, United Kingdom} \\
$^{3}$  {Department of Experimental Physics, Maynooth University, Maynooth Co. Kildare, Ireland}\\
   \today
}

\begin{document}

\maketitle


{\bf Abstract:}  {In this note, we revisit the scaling relations among ``hatted critical exponents'' which were first derived by
Ralph Kenna, Des Johnston and Wolfhard Janke, and we propose an alternative derivation for some of them. For the scaling relation involving the behavior of the correlation function, we will propose an alternative form since we believe that the expression is erroneous in the work of Ralph and his collaborators.}

{\bf Keyword:} {Critical exponents; Logarithmic corrections; Scaling and renormalization; Scaling laws; Tricritical point; Finite-Size Scaling}

\section*{In memory of our friend Ralph Kenna}
This paper is dedicated to our friend Ralph Kenna who passed away on Oct. 26th 2023. Ralph was a close collaborator and a very good friend, he was the PhD co-advisor of one of the authors of the present paper (LM). Renowned specialist in the study of phase transitions through the partition function zeros, he developed this formalism in the difficult cases where critical behaviours are controlled beyond the dominant singularities, by logarithmic corrections.

 The present work was initiated on the occasion of the ComPhys23 workshop\footnote{\href{http://www.physik.uni-leipzig.de/~janke/CompPhys23/}{http://www.physik.uni-leipzig.de/~janke/CompPhys23/}} organized by W. Janke in Leipzig, and dedicated to the memory of Ralph. The opening talk was intended to highlight some of Ralph's most important contributions to statistical physics, notably the derivation of scaling laws between the exponents associated with logarithmic singularities in the vicinity of second-order phase transitions. Ralph was a well-known physicist in the statistical physics community and these new scaling laws were very successful in the field of critical phenomena. It was by revisiting these scaling laws and their derivation that we noted that one of these relations was incomplete and that the work undertaken by Ralph and his co-authors around twenty years ago merited to be completed. 

This paper could have been written by Des Johnston or by Wolfhard Janke, the co-authors of Ralph in this topic (and in many more works).

\section{Scaling relations and universal combinations of  amplitudes, a short primer}\label{sec1}

We assume that the reader is familiar with the notion of critical exponents that describe the singularities of various thermodynamic functions at the approach of a second-order phase transition. Otherwise, we can suggest to refer e.g. to the textbook of Kardar~\cite{Kardar}.

Standard scaling relations among the universal critical exponents  are the following,
\bea
2\beta+\gamma &=& 2-\alpha,\label{eq1}\\
\beta(\delta -1) &=& \gamma,\label{eq2}\\
\nu(2-\eta) &=& \gamma,\label{eq3}\\
2-\alpha &=& \nu d.\label{eq4}
\eea
They are very useful, not only to obtain the values of all 6 fundamental critical exponents within a universality class from the knowledge of 2 of them but also because they allow for the definition of other universal quantities, written as specific combinations of critical amplitudes. Let us show how this works.
For that purpose, we first define the amplitudes as they enter the expressions of the leading singular behaviours of thermodynamic quantities in the vicinity of a second-order phase transition:
\bea
&&\hbox{Specific heat:} \quad C_\pm(t,0)\simeq A_\pm |t|^{-\alpha},\label{eq5} \\
&&\hbox{Low temperature magnetization:} \quad m_-(t,0)\simeq  B_- (-t)^{\beta},\label{eq6} \\
&&\hbox{Susceptibility:} \quad \chi_\pm(t,0)\simeq  \Gamma_\pm |t|^{-\gamma},\label{eq7} \\
&&\hbox{Critical temperature magnetization:} \quad m_c(0,h) \simeq B_c|h|^{1/\delta}, \label{eq8}\\
&&\hbox{Correlation length:} \quad  \xi_\pm(t,0) \simeq  \xi_{0\pm} |t|^{-\nu}.\label{eq9}
\eea
Here, the two arguments of the functions at the l.h.s. are respectively $t=(T-T_c)/T_c$ and $h=H/T_c$, 
and the indices $\pm$ specify the high ($t>0$) and low  ($t<0$) temperature phases, meaning that the field is zero, and the index $c$ on the contrary implies the field behaviour at the critical temperature.
The symbol $\simeq$ stands for the leading singularity (i.e. the most singular part, since there could be regular contributions to the thermodynamic quantities, power law corrections to scaling, and multiplicative logarithmic corrections, all these being omitted in Eqs.~(\ref{eq5})-(\ref{eq9})). 

We can also define the singular part of the  free energy density in zero field:
\be
f_\pm^{\rm sing}(t,0)\simeq F_\pm |t|^{2-\alpha},
\ee
and, since the specific heat is the second derivative of $f^{\rm sing}$ w.r.t $t$, $F_\pm$ is not independent, since this requires $A_\pm=(1-\alpha)(2-\alpha)F_\pm$. The Lee-Yang edge is another quantity of interest in critical phenomena and we define
\be
h_\pm^{\rm LY}(t)\simeq h_{0\pm}|t|^\Delta
\ee
with the so-called gap exponent $\Delta =\beta+\gamma=\beta\delta$. 

Universality is the observation that some quantities only depend on very general properties, like space dimensionality. The critical exponents are such universal quantities, but the amplitudes are not, although some combinations among them have the property of universality. To make it clear, let us write Widom's scaling assumption, i.e. the fact that the singular part of the free energy density is a homogeneous function of the scaling fields,
\be
f^{\rm sing}_\pm(t,h)=b^{-d}{\mathscr F}_\pm(\kappa_t b^{y_t}t,\kappa_h b^{y_h}h), \label{eq-scalingassumptionnolog}
\ee
where ${\mathscr F}_\pm(x(b),y(b))$ is a universal scaling function of its arguments, $y_t$ and $y_h$ are the RG dimensions of the relevant fields $t$ and $h$, and $\kappa_t$ and $\kappa_h$ are non universal metric factors which would
differ, say on the square lattice and the triangular lattice in $2d$. The amplitudes defined above depend on these metric factors, and this is why they are not universal. E.g. from $C_\pm(t,0)=\frac{\partial^2f^{\rm sing}_\pm(t,0)}{\partial t^2}$, setting $b=|t|^{-1/y_t}$ in the scaling form (\ref{eq-scalingassumptionnolog}), one obtains
\be C_\pm(t,0)=
 \kappa_t^{2+(d-2y_t)/y_t} |t|^{(d-2y_t)/y_t}\Bigl(\frac{\partial^2{\mathscr F}_\pm(x,0)}{\partial x^2}\Bigr)_{x=1}.\ee
This identifies the exponent \be \alpha=(2y_t-d)/y_t\ee and the amplitude \be A_\pm=\kappa_t^{2-\alpha}\Bigl(\frac{\partial^2{\mathscr F}_\pm(x,0)}{\partial x^2}\Bigr)_{x=1}.\ee 

The other exponents are similarly defined in terms of $y_t$ and $y_h$ by very famous relations that we do not repeat here and the other amplitudes depend on the metric factors as $B_-\sim \kappa_h\kappa_t^\beta$, $\Gamma_\pm\sim \kappa_h^2\kappa_t^{-\gamma}$, $B_c\sim \kappa_h^{1+1/\delta}$.

Simple ratios are immediately defined from the fact that the approach to criticality from above and from below is described by the same exponent for a given quantity (except for the magnetization, obviously). For example in the case of the specific heat  $C_+(|t|)\simeq A_+ |t|^{-\alpha}$ and $C_-(-|t|)\simeq A_- |t|^{-\alpha'}$, where $\alpha'=\alpha$. It follows  that the metric factors cancel in the ratio and
\be
R_C(|t|)=\frac{C_+(|t|)}{C_-(-|t|)}=\frac{A_+ |t|^{-\alpha}}{A_- |t|^{-\alpha'}} \to R_C=\frac{A_+}{A_-}
\ee
is thus universal. The limit corresponds to the approach to criticality ($|t|\to 0$ here) since the combination  $R_C(|t|)$ can be temperature-dependent due to the possible presence of different values for the amplitudes of the corrections to scaling which has not been taken into account in Eq.~(\ref{eq5}) to (\ref{eq9}). In the same manner, one defines the universal ratios
\be
R_\chi=\frac{\Gamma_+}{\Gamma_-},\quad
R_\xi=\frac{\xi_{0+}}{\xi_{0-}}.
\ee

The scaling relations are other examples of relations that allow the definition of new combinations. For example the ratio $m_-^2/\chi_\pm$ eliminates $\kappa_h$, and $\kappa_t$ is then eliminated,  thanks to Eq.~(\ref{eq1}) if we further divide by $C_\pm$. There is still an unwanted $|t|^2$ dependence that needs to be simplified and for that purpose
we  consider the quantity
\be
R_\pm^{(1)}(t) = \frac{m_-^2(-|t|)}{C_\pm(t)\chi_\pm(t) |t|^2}=
\frac{B_-^2}{C_\pm \Gamma_\pm} |t|^{2\beta+\gamma+\alpha-2} \to \frac{B_-^2}{C_\pm \Gamma_\pm}.
\ee
Thanks to Eq.~(\ref{eq1}), the fact that all metric factors cancel out in this latter quantity makes the combination ${B_-^2}/{C_\pm \Gamma_\pm}$ universal.
 Proceeding the same way, Eq.~(\ref{eq2}) and Eq.~(\ref{eq3}) suggest to contemplate the expressions
 \bea
 R_\pm^{(2)}(t) &=&\left. \chi_\pm(t) m_-^{\delta-1}(-|t|) m_c^{-\delta} (h)h\right|_{h=h_\pm^{\rm LY}(|t|)}
 \nnb\\
&&
 =\Gamma_\pm B_-^{\delta-1} B_c^{-\delta} |t|^{-\gamma+\beta(\delta-1)}  \to 
 \Gamma_\pm B_-^{\delta-1} B_c^{-\delta},\\
  R_\pm^{(3)}(t)&=&\frac{ \chi_\pm(t) } { \xi_\pm^{2-\eta}  (t) }=
 \frac{ \Gamma_\pm }{ \xi_{0\pm}^{2-\eta} } |t|^{-\gamma+\nu(2-\eta) }
  \to   \frac{\Gamma_\pm}{\xi_{0\pm}^{2-\eta}},
 \eea
 that reach their respective universal values.
Eventually, Eq.~(\ref{eq4}) leads to consider the following combination
\be
 R_\pm^{(4)}(t) = \xi_\pm^d(t) f_\pm^{\rm sing}(t)=
 \xi_{0\pm}^d F_\pm |t|^{-d\nu+2-\alpha} \to  \xi_{0\pm}^d F_\pm 
 \label{eq-ratio4}
\ee
as universal also.

\section{From the universal combinations of amplitudes to scaling laws among hatted exponents}\label{sec2}

Having the universal combinations of amplitudes at hand, we consider now the case where the critical behaviour is described, besides the leading singularities, by multiplicative logarithmic corrections.  This may happen for example for a system at its upper critical dimension $d_{\rm uc}$, or in the case of the $2d$ $4-$states Potts model, or the $2d$ disordered Ising model as well. Many examples can be found in Ref. \cite{RK1, RK2, RK3}.

Let us first remind the standard definitions of some exponent
combinations which will occur below: $\alpha_c=\alpha/\beta\delta$, $\gamma_c=\gamma/\beta\delta$,
$\nu_c=\nu/\beta\delta$, $\epsilon_c=(1-\alpha)/\beta\delta$. The logarithmic corrections can appear either in the approach to the critical temperature when the magnetic field is fixed at zero, or on the other hand right at $T_c$, when the magnetic field approaches zero:

\begin{eqnarray}
& h=0,\  t\to 0^\pm,  \\
&& f^{\rm sing}_\pm( t,0)\simeq F_\pm| t|^{2-\alpha}(-\ln| t|)^{\hat\alpha}, \label{eq-flog}\\
 &&m_-( t,0)\simeq B_-| t|^{\beta}(-\ln| t|)^{\hat\beta}, \\
 &&e_\pm( t,0)\simeq \frac{A_\pm}{(1-\alpha)}| t|^{1-\alpha}(-\ln| t|)^{\hat\alpha},  \\
&& \chi_\pm( t,0)\simeq \Gamma_\pm| t|^{-\gamma}(-\ln| t|)^{\hat\gamma},  \\
&& C_\pm( t,0)\simeq {A_\pm}| t|^{-\alpha}(-\ln| t|)^{\hat\alpha},  \label{eq-scalingClogt}\\
& &\xi_\pm( t,0)\simeq \xi_{0\pm}| t|^{-\nu}(-\ln| t|)^{\hat\nu},\label{eq-scalingXilogt}
\end{eqnarray}

\begin{eqnarray}
&  t=0,\ h\to 0^\pm, \\
&& f_c^{\rm sing}(0,h)\simeq F_c|h|^{1+1/\delta}(-\ln|h|)^{\hat\delta_c}, \label{eq-flogh}\\
& & m_c(0,h)\simeq B_c|h|^{1/\delta}(-\ln|h|)^{\hat\delta_c}, \label{eq-deltaLongAgo} \\
& & e_c(0,h)\simeq E_c|h|^{\epsilon_c}(-\ln|h|)^{\hat\epsilon_c}, \\
&& \chi_c(0,h)\simeq \Gamma_c|h|^{-\gamma_c}(-\ln|h|)^{\hat\delta_c}, \\
 & & C_c(0,h)\simeq \frac{A_c}{\alpha_c}|h|^{-\alpha_c}(-\ln|h|)^{\hat\alpha_c}, \\
& & \xi_c(0,h)\simeq \xi_c|h|^{-\nu_c}(-\ln|h|)^{\hat\nu_c}.
\end{eqnarray}
We can also define at criticality $t=h=0$ the logarithmic correction of the correlation function, defining the exponents $\hat\eta$ that will play an essential role in the following of this paper:
\be
G(0,0,|\vac r|)\simeq g_{0} |\vac r|^{-(d-2+\eta)}(\ln|\vac r|)^{\hat\eta}.
\ee
We mostly use the notations of
Refs.~\cite{PrivmanHohenbergAharony91,RK3}, with quantities (amplitudes and exponents) at the critical temperature defined with the subscript $c$, except for $\delta$ in (\ref{eq-deltaLongAgo}) which is standard according to the terminology fixed by Fisher long ago~\cite{FisherLongAgo}.

Ralph Kenna and his co-workers, Des Johnston and Wolfhard Janke, have established a series of scaling relations  \cite{RK1, RK2, RK3} among ``hatted exponents'', as Ralph was used to call them.
Their approach was based on the zeros of the partition function, either the Lee-Yang zeros (in complex magnetic field) or the Fisher zeros (in complex temperature).

Here, we offer an alternative derivation of most of these scaling laws, probably simpler in its approach.
Universality assumes that the previous ratios of amplitudes are still universal when multiplicative logarithmic corrections are present, i.e. 
\be
R_\pm^{(1)}(t) = \frac{m_-^2(-|t|)}{C_\pm(|t|)\chi_\pm(|t|) |t|^2}=
\frac{B_-^2}{C_\pm \Gamma_\pm} 
(-\ln|t|)^{2\hat \beta-\hat\alpha-\hat\gamma}.
\ee
The fact that this quantity {\em must} tend to ${B_-^2}/{C_\pm \Gamma_\pm}$ now demands that 
\be
2\hat \beta=\hat\alpha+\hat\gamma.\label{eq-RK1}
\ee
This is the first of Ralph and coworkers' scaling relations. Using the same method, the second ratio easily leads to a second relation, 
\be
\hat\gamma+\hat\beta(\delta -1)-\delta\hat\delta=0.\label{eq-RK2}
\ee
The amplitude of the Lee-Yang edge, $h_{0\pm}$, has a non trivial dependence with the metric factors, $h_{0\pm}\sim \kappa_t^{\beta\delta}\kappa_h^{-1}$. This can be retrieved from the scaling relation $\Delta=\beta\delta$, and the universality of the ratio
\be
R^{(5)}(t)=
\frac{m_c(h^{\rm LY}_\pm(t))}{m_-(t)}=\frac{B_ch_{0\pm}^{1/\delta}}{B_-}
|t|^{\Delta/\delta-\beta}(-\ln|t|)^{\hat\Delta/\delta+\hat\delta-\hat\beta}\to \frac{B_ch_{0\pm}^{1/\delta}}{B_-}
\ee
requires that
\be
\hat\Delta=(\hat\beta-\hat\delta)\delta. \label{eq-RKDelta}
\ee
 The scaling relations (\ref{eq-RK1}) and (\ref{eq-RK2}) were first derived in Ref.~\cite{RK1}. Instead of (\ref{eq-RKDelta}), Ralph and his co-workers had $\hat\Delta=\hat\beta-\hat\gamma$ which is recovered here using  (\ref{eq-RK2}) and (\ref{eq-RKDelta}).
 
 In the same paper, they also derived 
 \be \hat\koppa=\hat\nu+\nu\hat\alpha/(2-\alpha).\label{eq-hatkoppaRK}\ee 
 This is an analogue of the hyperscaling relation for logarithmic relations. A new {\em pseudo-critical exponent}~\cite{RK1,RK2,KBCMP13} appears there, $\hat\koppa$, that describes the finite-size scaling (FSS) of the correlation length,
 \be
 \xi_L\simeq L(\ln L)^{\hat\smallkoppa}.\label{eq-FSSXiLog}
 \ee
 This behaviour is encoded in the scaling hypothesis for the correlation length, appropriately extended to account for the logarithmic correction:
 \be
 \xi_\pm(t,h,L^{-1})=b(\ln b)^{\hat\smallkoppa}{\mathscr X}(\kappa_t b^{y_t}(\ln b)^{\hat y_t}t,\kappa_h b^{y_h}(\ln b)^{\hat y_h}h,bL^{-1}).\label{eq-scalingassumptionxilog}
 \ee
 Like $y_t$ (resp. $y_h$) is the RG eigenvalue associated with the scaling field $t$ (resp. $h$), we denote ${\hat y}_t$ (resp. ${\hat y}_h$) the corresponding exponent of the logarithmic correction.
For the sake of clarity, we will later denote the rescaled variables as
$x(b)=\kappa_t b^{y_t}(\ln b)^{\hat y_t}t$, $y(b)=\kappa_h b^{y_h}(\ln b)^{\hat y_h}h$, and $z(b)=bL^{-1}$.
 Eq.~(\ref{eq-FSSXiLog}) follows from the choice $b=L$
at criticality $t=h=0$ in  (\ref{eq-scalingassumptionxilog}). The same scaling form is used in the thermodynamic limit $L\to\infty$, setting $x=1$. This requires iterations
 \bea
 b&=&(\kappa_t |t|)^{-1/y_t} (\ln b)^{-\hat y_t/y_t}\nnb\\ &\simeq& 
 (\kappa_t |t|)^{-1/y_t} (-\ln |t|)^{-\hat y_t/y_t}
 \Bigl(
 1+\frac{\ln(-\ln|t|)}{(-\ln |t|)}
  +
 \ \hbox{higher order correction}
 \Bigr).\nonumber\\ \label{eq-highorders}
 \eea 
 Inserted in the expression of  the correlation length leads to leading order to
\be
\xi_\pm(t,0,0)\simeq |t|^{-1/y_t}(-\ln|t|)^{\hat\smallkoppa-\hat y_t/y_t}
{\mathscr X}(1,0,0)
\ee
and requires the usual relation $\nu=1/y_t$, but also, to conform to (\ref{eq-scalingXilogt}),
\be
\hat\koppa=\hat\nu+\nu\hat y_t. \label{eq-hqtkoppa1}
\ee 
We now show that this agrees with Ralph's scaling relation (\ref{eq-hatkoppaRK}). For that purpose, we use the compatibility with the extension of the phenomenological Widom 
 scaling assumption for the free energy density (\ref{eq-scalingassumptionnolog}) to the presence of logarithmic corrections,   written as far as we know for the first time by Ralph Kenna in Ref.~\cite{RKNPB}, 
  \be
f^{\rm sing}_\pm(t,h,L^{-1})=b^{-d}{\mathscr F}_\pm(\kappa_t b^{y_t}(\ln b)^{\hat y_t}t,\kappa_h b^{y_h}(\ln b)^{\hat y_h}h,bL^{-1}) \label{eq-scalingassumptionlog}.
\ee
The second derivative w.r.t. $t$ is the specific heat, and the choice $x=1$ at $h=L^{-1}=0$ then leads (using $\alpha=(2y_t-d)/y_t$) to 
\be
C_\pm(t,0,0)\simeq |t|^{-\alpha}(-\ln|t|)^{2\hat y_t-\alpha\hat y_t}{\mathscr C}_\pm(1,0,0),
\ee (from now on, we always limit (\ref{eq-highorders}) to leading logarithmic order), hence, from (\ref{eq-scalingClogt}),
\be
\hat \alpha=(2-\alpha)\hat y_t
\ee
which completes the proof.

\section{Solving a disagreement with our friends}

 In Ref.~\cite{RK3}, Ralph Kenna has given a complete account of these, and many more scaling relations among hatted exponents. This is not our purpose here to be exhaustive, but rather to show alternative derivations, or to complete what Ralph and coworkers didn't do. With this perspective in mind, Eqs.~(\ref{eq-scalingassumptionxilog}), (\ref{eq-scalingassumptionlog}) and an analogous homogeneity form for the correlation function (discussed later) offers an option to proceed as we show now. 

In Ref.~\cite{RK2}, two other scaling relations between hatted exponents were derived:
\bea
&&\hat\alpha=d(\hat\koppa-\hat\nu)\quad\hbox{or}\quad \hat\alpha=1+d(\hat\koppa-\hat\nu),\label{eq-hyperscalinglog}\\
&&\hat\eta=\hat\gamma-\hat\nu(2-\eta).\label{eq-RKhat-eta}
\eea
Concerning the first relation (\ref{eq-hyperscalinglog}), the second formula is valid in such circumstances where the model has $\alpha=0$ and an impact angle $\phi\not=\pi/4$ for the Fisher zeros in the complex plane (this is the case for the pure two-dimensional Ising model). We will not consider this case, but rather the more general case of the first formula. It can be rederived by careful use 
of the ratio 4 (Eq.~(\ref{eq-ratio4})) in section \ref{sec1}, and  even requires to use of FSS of the correlation length.
 From $\xi_\pm(t)\simeq |t|^{-\nu}(-\ln|t|)^{\hat \nu}$, we first reverse
 to $|t|\simeq \xi_\pm^{-1/\nu}(t)(\ln\xi_\pm(t))^{\hat \nu/\nu}$. 
  This expression is then incorporated into (\ref{eq-flog}) to get  
  \be f^{\rm sing}_\pm (t)\simeq\xi_\pm^{-d}(t)(\ln\xi_\pm(t))^{d\hat\nu+\hat\alpha}\label{eq-fvsxi}\ee 
  i.e. a modified version of (\ref{eq-ratio4}):
  \be
 R_\pm^{(4bis)}(t) = \frac{\xi_\pm^d(t) f_\pm^{\rm sing}(t)}{(\ln\xi_\pm(t))^{d\hat\nu+\hat\alpha}}\to  \xi_{0\pm}^d F_\pm .
 \label{eq-ratio4modified}
\ee

Now, inserting (\ref{eq-FSSXiLog}) into (\ref{eq-fvsxi}) leads to the FSS behaviour of the 
 free energy density  at criticality,
 \be
 f^{\rm sing}_L(0)\simeq L^{-d}(\ln L)^{-d\hat\smallkoppa}(\ln L)^{\hat\alpha+d\hat\nu},
 \ee
 and compatibility with (\ref{eq-scalingassumptionlog}) at $t=h=0$, $b=L$
 then demands 
 \be
 \hat\alpha=d\hat\smallkoppa-d\hat\nu.\label{eq-RK4}
 \ee
 which is Kenna and coworkers' relation.
 
 The same derivation can be done for the magnetic sector, considering the approach to criticality at $T_c$ for $h\to 0$ and yields the scaling relation
  \be
 \hat\delta=d\hat\smallkoppa-d\hat\nu_c.\label{eq-RK4bis}
 \ee

Concerning the last Eq.~(\ref{eq-RKhat-eta}), we believe that this relation is incomplete. Applied to the $4-$state Potts model in two dimensions~\cite{Nauenberg,Cardy2,2dPM,SBBFirst}, which has  $\hat\eta=-\frac 18$, $\hat\gamma=\frac 34$, $\hat\nu=\frac 12$ and $\eta=\frac 14$, Eq.~(\ref{eq-RKhat-eta}) is fulfilled. We believe that this is because there $\hat\koppa=0$, and that an additional term $\hat\koppa\gamma/\nu$ is missing in the general case.
A test is provided in the case of the Ising model in 4 dimensions which has $\hat\koppa=\frac 14$ (models at their upper critical dimensions have $\hat\koppa=1/d_{\rm uc}$~\cite{KBCMP13,KBFisherScaling,JJRL17,KBPrevioulsyUnpubmlished}). There, Ralph and his co-authors had anticipated that 
$\hat\eta=0$ for Eq.~(\ref{eq-RKhat-eta}) to work ($\hat\gamma=\frac 13$, $\hat\nu=\frac 16$ and $\eta=0$ for the Ising model in 4 dimensions), but according to Luijten~\cite{Lthesis}, $\hat\eta=\frac 12$ instead, a result that is in contradiction with Eq.~(\ref{eq-RKhat-eta}).

Let us examine the problem in more detail. In Ref.~\cite{RK2}, the authors have questioned the relation between the correlation function and the square of the magnetization when the system decorrelates, i.e. for $|\vac r|\to\infty$:
\be
G(t,h,L^{-1}\to\infty,|\vac r|\to\infty)\to m^2(t,h,L^{-1}\to\infty).
\ee
On the contrary, we assume that there is no reason why this would not be valid, so we start from the homogeneity of the (spin-spin) correlation function with logarithmic corrections as 
\be
G(t,h,L^{-1},|\vac r|)= b^{-(d-2+\eta)}(\ln b)^{\hat\eta}G(x(b),y(b),z(b),|\vac r|/b),\label{eq-homogG}
\ee
Setting $x=1$, $y=0$ and the thermodynamic limit $z=0$ leads to the following temperature behaviour when $|\vac r|\to\infty$
\be
G(t,0,0,\infty)\simeq |t|^{\frac{d-2+\eta}{y_t}}(-\ln|t|)^{\frac{d-2+\eta}{y_t}\hat y_t+\hat\eta}G(1,0,0,\infty)=m^2(t,0,0).
\ee
This requires the usual relation $2\beta=\frac{d-2+\eta}{y_t}$ and
\be
\hat \eta=2(\hat\beta-\beta\hat y_t).\label{eq-hatetabeta}
\ee
The two examples given above are test grounds.  For the $4-$state Potts model in two dimensions, we extract immediately $\hat\eta=-\frac 18$ which is correct. For the $4d$ Ising model on the other hand, we obtain $\hat\eta=\frac 12$, in agreement with Luijten's result~\cite{Lthesis}, later verified numerically in Ref.~\cite{Deng}, but we are here in contradiction with the prediction of Refs.~\cite{RK2,RK3}.

Since the question is of importance, we want to consider it from other perspectives also. The correlation function is linked to the susceptibility via the fluctuation-dissipation relation: 
\be
\chi(0,0,L^{-1})=\int_0^Ld^d r\ \!|\vac r|^{-(d-2+\eta)}(\ln|\vac r|)^{\hat\eta}G(0,0,|\vac r|L^{-1},1).
\ee
Setting $\rho=|\vac r|/L$, we have
\bea
\chi(0,0,L^{-1})&=&L^{2-\eta}(\ln L)^{\hat\eta}\int_0^1d^d\rho\ \! \rho^{-(d-2+\eta)}\Bigl(1+\frac{\ln \rho}{\ln L}\Bigr)^{\hat\eta}G(0,0,\rho,1)\nonumber\\
&\simeq& L^{2-\eta}(\ln L)^{\hat\eta},\label{eq-FSSFlucDiss}
\eea
and, since the susceptibility obeys, via the second derivative of (\ref{eq-scalingassumptionlog}), to
\be
\chi(t,h,L^{-1})=\kappa_h^2 b^{-d+2y_h}(\ln b)^{2\hat y_h}{\mathscr Y}(x(b),y(b),z(b)),
\label{eq-chihomog}
\ee
its FSS compared to (\ref{eq-FSSFlucDiss}) demands that $2-\eta=2y_d-d=\gamma/\nu$ and
\be
\hat\eta=2\hat y_h.\label{Eq-etahatbyyhhat}
\ee
Again, this confirms $\hat\eta=\frac 12$ for the $4d$ Ising model.

The fluctuation-dissipation theorem has also been used in Refs.~\cite{RK2,RK3} in the form 
\be\chi_\infty (t)\sim \xi_\infty^{2-\eta}(t)(\ln\xi_\infty(t))^{\hat\eta},\label{FSSFlucDissRK}\ee 
from where Eq.~(\ref{eq-RKhat-eta}) follows,
so there is still some difficulty hidden to solve our disagreement.
Let us set $b=\xi_\infty(t)$ in (\ref{eq-chihomog}):
\be
\chi_\infty(t)=\xi_\infty^{-d+2y_h}(t) (\ln\xi_\infty (t))^{\hat\eta}{\mathscr Y}(x(\xi_\infty(t)),0)
\ee
where the variable $x$ in the scaling function ${\mathscr Y}$ is evaluated at $\xi_\infty(t)$ to give
\be
x(\xi_\infty(t))=\xi_\infty(t)^{y_t}(\ln\xi_\infty(t))^{\hat y_t}|t|=(-\ln|t|)^{\hat\smallkoppa/\nu}.
\ee
The scaling function must have the behaviour ${\mathscr Y}(x,0)\sim x^{-\gamma}$ when $|t|\to 0$ to recover the temperature singularity of the susceptibility $\chi_\infty(t,0)\sim|t|^{-\gamma}(-\ln|t|)^{\hat\gamma}$. It follows that instead of (\ref{FSSFlucDissRK}), one has
\be\chi_\infty (t)\sim \xi_\infty^{2-\eta}(t)(\ln\xi_\infty(t))^{\hat\eta-\gamma\hat\smallkoppa/\nu}
\label{FSSFlucDissRKmod}\ee 
and instead of
Eq.~(\ref{eq-RKhat-eta}), one has a third form for the exponent $\hat\eta$:
\be
\hat\eta=\hat\gamma-\hat\nu(2-\eta)+\gamma\frac{\hat\smallkoppa}{\nu},\label{eq-RKhat-etamod}
\ee
again compatible with $\hat\eta=\frac 12$ for the $4d$ Ising model.
This suggests to use, instead of  the ratio
$ R_\pm^{(3)}(t)$, the modified version
\bea
  R_\pm^{(3{\rm bis})}(t)&=&\frac{ \chi_\pm(t) } {\xi_\pm^{2-\eta}(t) (\ln \xi_\pm(t))^{\hat\eta-\gamma\hat\smallkoppa/\nu}}\nnb\\
  &=&\frac{\Gamma_\pm}{\xi_{0\pm}^{2-\eta}}(-\ln|t|)^{\hat\gamma-\hat\eta-\gamma\hat\smallkoppa/\nu-\hat\nu(2-\eta)} \to   \frac{\Gamma_\pm}{\xi_{0\pm}^{2-\eta}}
  \eea
  which, again, is universal. The 4 standard scaling laws and the corresponding 4 hatted scaling laws are listed in table~\ref{tableRatios}.

 \begin{table}[ht]
 \begin{center}
  \begin{tabular}{c|c|c}
  \hline
\hbox{\vrule depth 10pt height 13pt width 0pt}   Ratio & scaling relation & hatted scaling relation \\
    \hline
\hbox{\vrule depth 10pt height 14pt width 0pt} $\frac{m_-^2(-|t|)}{C_\pm(|t|)\chi_\pm(|t|) |t|^2}$ & $2\beta+\gamma=2-\alpha  $ & $2\hat \beta=\hat\alpha+\hat\gamma$ \\
\hbox{\vrule depth 15pt height 13pt width 0pt}
 $\left. \frac{\chi_\pm(t) m_-^{\delta-1}(-|t|) h}{m_c^{\delta} (h)}\right|_{h=h_\pm^{\rm LY}(|t|)}$ & $ \beta(\delta -1) = \gamma$& $\hat\gamma+\hat\beta(\delta -1)-\delta\hat\delta=0$ \\
 \hbox{\vrule depth 10pt height 13pt width 0pt}
 $ \frac{ \chi_\pm(t) } {\xi_\pm^{2-\eta}(t) (\ln \xi_\pm(t))^{\hat\eta-\gamma\hat\smallkoppa/\nu}}$ & $\nu(2-\eta)=\gamma$ & $\hat\gamma-\hat\nu(2-\eta)
+\gamma\frac{\hat\smallkoppa}{\nu} =\hat\eta$ \\
\hbox{\vrule depth 10pt height 13pt width 0pt} $ \xi_\pm^d(t) f_\pm^{\rm sing}(t)$ & $2-\alpha=\nu d$ & $\hat\alpha=d\hat\koppa-d\hat\nu$ \\
   \hline
  \end{tabular}
  \caption{The main scaling laws and hatted scaling laws and the associated universal combinations of amplitudes.\label{tableRatios}}
   \end{center}
  \end{table}

So far so good, but the situation is not yet clear, since the case of percolation in 6 dimensions (its upper critical dimension) is maybe a counterexample.
With $\gamma=1$, $\nu=\frac 12$, $\eta=0$, and the values of the logarithmic correction exponents $\hat\gamma=\frac 27$, $\hat\nu=\frac 5{42}$ and of the pseudo-critical exponent $\hat\koppa=\frac 1{d_{\rm uc}}=\frac 16$, using Eq.~(\ref{eq-RKhat-etamod})
we predict $\hat\eta=\frac 8{21}$, while Kenna and coworkers predict $\hat\eta=\frac 1{21}$ from Eq.~(\ref{eq-RKhat-eta}). This latter result conforms to an analytic prediction from Ref.~\cite{Stenull}, but on the other hand, our value is supported by an FSS prediction by Ruiz-Lorenzo~\cite{Ruiz-Lorenzo}, $\chi_L L^{-2}\sim (\ln L)^{8/21}$. This disagreement demands further analysis.

 In Ref.~\cite{RK3}, Kenna has listed the values of the known hatted critical exponents for a series of models, and when $\hat\eta$ was not known, he has proposed the expected value from the use of Eq.~(\ref{eq-RKhat-eta}). An interesting model is missing from the list, the tricritical Ising model, which has logarithmic corrections at its upper critical dimension $d_{\rm uc}=3$ and has non-zero $\hat\koppa=1/3$. We will now analyse this universality class in more detail.

  \section{The tricritical Ising universality class in the Blume-Capel model at the upper critical dimension}
 
The  spin-$1$ Blume-Capel model~\cite{BC1,BC2} is a lattice model defined by the Hamiltonian 
\begin{equation}
    \mathcal{H}=-J\sum_{\langle i,j\rangle}\sigma_i \sigma_j + \Delta \sum_i \sigma_i^2 - H\sum_i \sigma_i ,
\end{equation}
where the spin variables $\sigma_i =-1, 0, +1$,  $J > 0$ denotes the ferromagnetic exchange interaction among nearest-neighbour sites ($\langle i,j \rangle$ indicates summation over nearest neighbours), and $\Delta$ is the crystal-field strength that controls the density of vacancies (the $\sigma_i = 0$ states can be viewed as vacancies in an ordinary $\sigma_i=\pm1$ Ising model)~\cite{Cardy}. For $\Delta = -\infty$, vacancies are suppressed from the partition function and the Hamiltonian reduces to that of the Ising ferromagnet. At $\Delta =0$ the second-order transition is in the pure Ising model universality class. When $\Delta$ increases from $0$, a perturbation theory shows that the transition temperature decreases along a line which remains of Ising-like second-order phase transition.
On the other hand, in the vicinity of $T=0$ the transition is first-order and persists first-order at small values of $T$ until it reaches the second-order line.
Right at the limit, there is a tricritical point characterized by specific values of $T_t$, $\Delta_t$. Tricriticality corresponds to the $\phi^6$ Landau expansion~\cite{SciPost} and the upper critical dimension is thus $d_{\rm uc}=3$, the case that we consider now.

 \begin{figure}[ht] \centering
\includegraphics[width=0.7\linewidth]{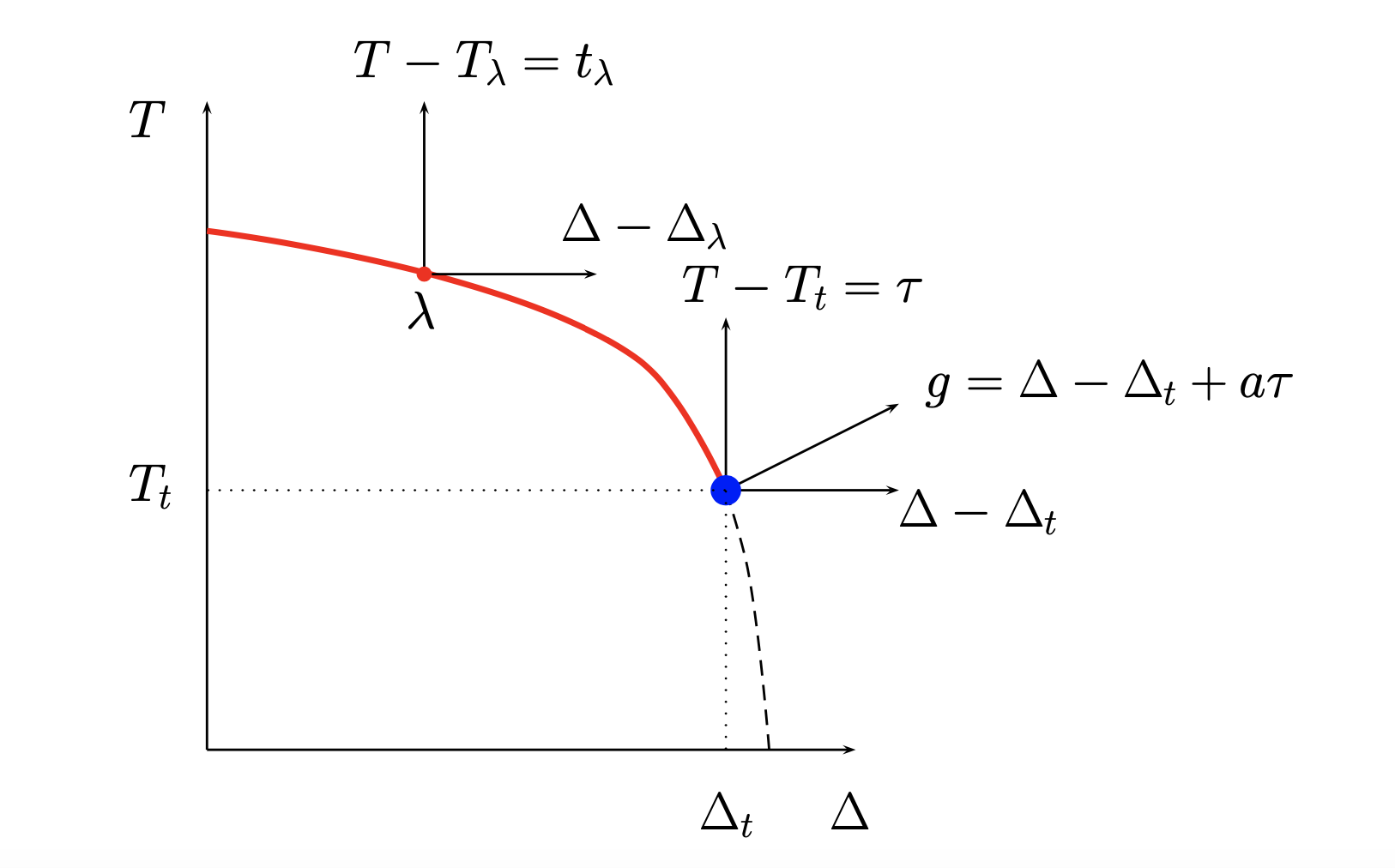} 
\caption{Typical phase diagram of the Blume-Capel model in the $(T,\Delta)$ plane. The $\lambda$ line is a line of second-order phase transition in the Ising model universality class that ends at a tricritical point of coordinates $(T_t,\Delta_t)$. The dashed line is a first-order transition line. The most singular even scaling field at the tricritical point is $g$, a linear combination of $\tau$ and of $\Delta-\Delta_t$.\label{Fig-PhaseDiag}
}\end{figure}

Using appropriate linear combinations of the physical parameters,  determined by the geometry of the phase diagram and the fact that they have to vanish at the tricritical point, the even scaling fields are $\tau=(T-T_t)/T_t$,  $g=(\Delta-\Delta_t)/T_t+a\tau$, and the odd scaling field is the usual magnetic field $h=H/T_t$. Lawrie and Sarbach~\cite{LS} have shown that the free energy density at the tricritical point in $3d$ reads, in terms of these scaling fields, as
 \begin{equation}
    f^{\rm sing}_{\rm tri}(\tau,g,h)=b^{-3} {\mathscr F}_{\pm} (b^{1}(\ln b)^{\frac{4}{15}}\tau, b^{2}(\ln b)^{\frac13}g(1-a^{-1}b^{-1} (\ln b)^{-\frac1{15}}\tau g),b^{\frac 52} (\ln b)^{\frac{1}{6}}h),
    \label{eq-sws}
\end{equation}
 where the non universal metric factors $\kappa_i$ have been omitted. Here, the subscript `tri' indicates that the expression is valid in the vicinity of the tricritical point where $\tau=g=h=0$. 
The logarithmic corrections are explicitly given in this expression. 
 
 The notations in Ref.~\cite{LS} necessitate to be adapted to be consistent with those that we have used until now. Leaving aside the logarithmic corrections for a while, let us write
  \begin{equation}
    f^{\rm sing}_{\rm tri}(\tau,g,h)=b^{-d} {\mathscr F}_{\pm} (b^{y_\tau}\tau, b^{y_g}g,b^{y_h}h).
    \label{eq-sws-simple}
\end{equation}
The dominant even scaling field is $g$, the usual singularities and critical exponents are therefore defined by their behaviours w.r.t. $g$ instead of $\tau$ which brings corrections to scaling due to crossover.  This means that the $t$ and $y_t$ of the three previous sections of this paper will now be replaced by $g$ and $y_g$.
At $h=0$ and $\tau=0$, setting $b=|g|^{-1/y_g}$ leads to  
 $  f^{\rm sing}_{\rm tri}(0,g,0)\sim|g|^{d/y_g}$. This is compatible with  $\nu=1/y_g$. The specific heat measures the total energy fluctuations. Its most singular part is defined by
 \bea
&& C(\tau,g,h)=\frac{\partial^2f^{\rm sing}_{\rm tri}}{\partial g^2}=b^{-d+2y_g}{\mathscr C}_\pm (b^{y_\tau}\tau, b^{y_g}g,b^{y_h}h)\sim |g|^{-\alpha}\quad\hbox{when}\ \tau,h=0\nonumber\\
 \eea
 with $\alpha=\frac{2y_g-d}{y_g}$.
This shows that the exponent $d/y_g$ of $f^{\rm sing}_{\rm tri}(0,g,0)$ is thus equal to the usual value $2-\alpha$
and the hyperscaling relation holds.

We can also define less singular exponents w.r.t. the scaling field $\tau$ (with tilde notation), e.g. $C(\tau,0,0)\sim|\tau|^{-\tilde\alpha}$. A similar analysis as above shows that   $f^{\rm sing}_{\rm tri}(\tau,0,0)\sim|\tau|^{d/y_\tau}$ with $d/y_\tau=(d/y_g)\phi=(2-\alpha)\phi$ with $\phi=y_g/y_\tau$ the crossover exponent. There is a caveat here since $d/y_\tau$ {\em is not equal to} $2-\tilde\alpha$ as one can find in the literature~\cite{LS}. Indeed, $\tilde\alpha=(2y_g-d)/y_\tau=\alpha\phi$, hence $2-\tilde\alpha=2-\alpha\phi\not=(2-\alpha)\phi$.
This is important to collect correct expressions, and this is done 
in table~\ref{tabLeadingSub}.

\begin{table}[h]
 {\footnotesize
\begin{tabular}{cc | cc | cc}
\hline
\multicolumn{2}{c}{\vrule depth 8pt height 12pt width 0pt leading even field, $g$}   &\multicolumn{2}{c}{subleading even field, $\tau$} & \multicolumn{2}{c}{leading odd field, $h$}  \\
\hline
\hbox{\vrule depth 10pt height 12pt width 0pt}   
$C(0,g,0)\sim |g|^{-\alpha}$ & $\alpha=\frac{2y_g-d}{y_g}$ & $C(\tau,0,0)\sim |\tau|^{-\tilde\alpha}$ & $\tilde\alpha=\alpha\phi$ \\
\hbox{\vrule depth 10pt height 8pt width 0pt}  $m(0,g,0)\sim |g|^{\beta}$ & $\beta=\frac{d-y_h}{y_g}$ & $m(\tau,0,0)\sim |\tau|^{\tilde\beta}$ & $\tilde\beta=\beta\phi$ & $m(0,0,h)\sim |h|^{1/\delta}$ & $\delta=\frac{y_h}{d-y_h}$\\
\hbox{\vrule depth 10pt height 8pt width 0pt} $ \chi(0,g,0)\sim |g|^{-\gamma}$ & $\gamma=\frac{2y_h-d}{y_g}$ & $ \chi(\tau,0,0)\sim |\tau|^{-\tilde\gamma}$ & $\tilde\gamma=\gamma\phi$ \\
\hline
\end{tabular}
}
\caption{Leading and subleading singularities for the most common physical quantities and the definitions of the associated exponents. Here the crossover exponent is $\phi=y_g/y_\tau$.\label{tabLeadingSub}}
\end{table}


This being said, we can now incorporate the logarithmic corrections in Eq.~(\ref{eq-sws-simple}) to get
  \begin{equation}
    f^{\rm sing}_{\rm tri}(\tau,g,h)=b^{-d} {\mathscr F}_{\pm} (b^{y_\tau}(\ln b)^{{\hat y}_\tau}\tau, b^{y_g}(\ln b)^{{\hat y}_g}g,b^{y_h}(\ln b)^{{\hat y}_h}h),
    \label{eq-sws-withlogs}
\end{equation}
and the comparison with Eq.~(\ref{eq-sws}) simply leads to
\bea
&&y_g=2,\quad y_h=\frac 52,\quad y_\tau=1,\quad\\
&&\hat y_g=\frac 13,\quad \hat y_h=\frac 16,\quad \hat y_\tau=\frac 4{15},\quad\hat\koppa=\frac 13.
\eea

 \begin{table}[h!] 
 {\footnotesize
\begin{tabular}{c | cccc | cccc}
\hline
 &
\multicolumn{4}{c}{\vrule depth 8pt height 12pt width 0pt  
  leading exponent}
  & 
 \multicolumn{4}{c}{\vrule depth 8pt height 12pt width 0pt  
  logarithmic correction exponent} \\
\hbox{\vrule depth 10pt height 13pt width 0pt} Quantity
& & IM4D & Tri. IM & Perco.&  
 & IM4D & Tri. IM & Perco.\\
\hline
\hbox{\vrule depth 10pt height 13pt width 0pt}   $C(t,0)$ &   $\alpha=\frac{2y_t-d}{y_t}$ & $0$ & $ \frac 12$ & $ -1$ & $ \hat\alpha=(2-\alpha)\hat y_t$  &$\frac 13$ & $\frac 12$ & $\frac 27$ \\
\hbox{\vrule depth 10pt height 13pt width 0pt}   $m_-(t,0)$ &   $\beta=\frac{d-y_h}{y_t}$ & $\frac12$ & $ \frac 14$ & $ 1$ & $ \hat\beta=\beta\hat y_t+\hat y_h$  &$\frac 13$ & $\frac 14$ & $\frac 27$ \\
\hbox{\vrule depth 10pt height 13pt width 0pt}   $\chi(t,0)$ &   $\gamma=\frac{2y_h-d}{y_t}$ & $1$ & $ 1$ & $ 1$ & $ \hat\gamma=2\hat y_h-\gamma\hat y_t$  &$\frac 13$ & $0$ & $\frac 27$ \\
\hbox{\vrule depth 10pt height 13pt width 0pt}     $m_c(0,h)$ &   $\frac1\delta=\frac{d-y_h}{y_h}$ & $\frac13$ & $\frac 15$ & $\frac 12$ & $ \hat\delta=\frac1\delta\hat y_h+\hat y_h$  &$\frac 13$ & $\frac 15$ & $\frac 27$ \\
\hbox{\vrule depth 10pt height 13pt width 0pt}     $h_{\rm LY}(t,0)$ &   $\Delta=\frac{y_h}{y_t}$ & $\frac 32$ & $\frac 54$ & $2$ & $ \hat\Delta=\Delta\hat y_t-\hat y_h$  &$0$ & $\frac 14$ & $0$ \\
\hbox{\vrule depth 10pt height 13pt width 0pt}   $\xi(t,0)$ &   $\nu=\frac{1}{y_t}$ & $\frac 12$ & $\frac 12$ & $ \frac 12$ & $ \hat\nu=\hat\koppa-\nu\hat y_t$  &$\frac 16$ & $\frac 16$ & $\frac 5{42}$ \\
\hbox{\vrule depth 10pt height 13pt width 0pt}   $G(0,0,|\vac r|)$ &   $\eta=d-2y_h+2$ & $0$ & $0$ & $0$ & $ \hat\eta=2\hat y_h$  &$\frac 12$ & $\frac 13$ & $\frac 8{21}$ \\
\hline
\end{tabular}
}
\caption{Leading and logarithmic correction exponents for the most common physical quantities.\label{tabSummary}}
\end{table}

We can now deduce the values of the standard critical exponents and the associated logarithmic corrections exponents. They are listed in Table~\ref{tabSummary} for three universality classes which all have non-zero values of the pseudo-critical exponent $\hat\koppa$, the Ising model in four dimensions, the tricritical Ising model in three dimensions, and the problem of percolation in six dimensions.

 In this table, the first six lines are not controversial. The seventh line presents the correlation function correction exponent $\hat\eta$ which follows from our scaling law, in any of the forms given in Eqs.~(\ref{eq-hatetabeta}), (\ref{Eq-etahatbyyhhat}) or (\ref{eq-RKhat-etamod}). These three expressions are mutually consistent, but they differ from Eq.~(\ref{eq-RKhat-eta}) used by Kenna and his co-authors. This latter formula would respectively predict for the three universality classes the values $0$, $-\frac 13$
 and $\frac 1{21}$. We have seen that the first value, $0$, is falsified in the $4d$ IM case by Refs.~\cite{Lthesis,Deng}, but the results of Ref.~\cite{Stenull} invalidates our third value
 $\frac 8{21}$, the last entry of in table~\ref{tabSummary}
while Ref.~\cite{Ruiz-Lorenzo} on the contrary supports this value.

The case of the tricritical Ising model in three dimensions appears crucial and we have to provide numerical results in support of our result.
The numerical computation of the correlation function is known to be a very delicate problem and we will approach the value of the exponent $\hat\eta$ differently, using FSS. Another delicate aspect is the also well-known fact that extracting logarithmic corrections in the vicinity of a critical point can be extremely difficult~\cite{BBS}.
Recently, it was found that very accurate results can be obtained numerically in the Blume-Capel model with relatively small system sizes~\cite{LMetal} via the analysis of the zeros of the partition function, and in particular the Lee-Yang zeros~\cite{LY,LY2}.
The Lee-Yang zeros are connected to the susceptibility~\cite{NPB} via
\be
\chi(g,0,L^{-1})\simeq L^{-d} \sum_{j=1}^{L^d} h_j^{-2}(g,0,L^{-1}),
\ee
where $j$ labels the zeros
 in the upper half complex plane are indexed in order of increasing distance from the critical point. 
The sum is dominated by the lowest zero, the Lee-Yang edge $h^{\rm LY}$, and at the tricritical point, the FSS of the susceptibility is thus linked to that of $h^{\rm LY}$:
\be
\chi(0,0,L^{-1})\simeq L^{-d} h^{\rm LY}(0,0,L^{-1})^{-2}\simeq L^{2-\eta}(\ln L)^{\hat\eta}.\label{Eq-chi-h}
\ee
In the presence of logarithmic corrections to the scaling form of the Lee-Yang edge obeys
\be
h^{\rm LY}(g,h)=b^{- y_h}(\ln b)^{-\hat y_h}{\mathscr H}_\pm
 (b^{y_\tau}(\ln b)^{{\hat y}_\tau}\tau,b^{y_g}(\ln b)^{{\hat y}_g}g,b^{y_h}(\ln b)^{{\hat y}_h}h)
\ee
{
compatible with the behaviour in terms of the thermal scaling field $g$,
as it can be shown using the scaling laws of Table~\ref{tabSummary},
$h^{\rm LY}(g,0)\sim |g|^\Delta (-\ln|g|)^{\hat\Delta}$.
}
If one sits exactly at the tricritical point, $\tau=g=h=0$, we can extract the FSS behaviour of the zeros by setting $b=L$,
\be
h^{\rm LY}(0,0,L^{-1})\simeq L^{-y_h}(\ln L)^{-\hat y_h}
\ee
and it follows that we expect 
\be
h^{\rm LY}(0,0,L^{-1})\simeq L^{(\eta-2-d)/2}(\ln L)^{-\hat \eta/2}
\ee
which agrees with Eq.~(\ref{Eq-chi-h}).

\begin{figure}[ht] \centering
\includegraphics[width=0.8\linewidth]{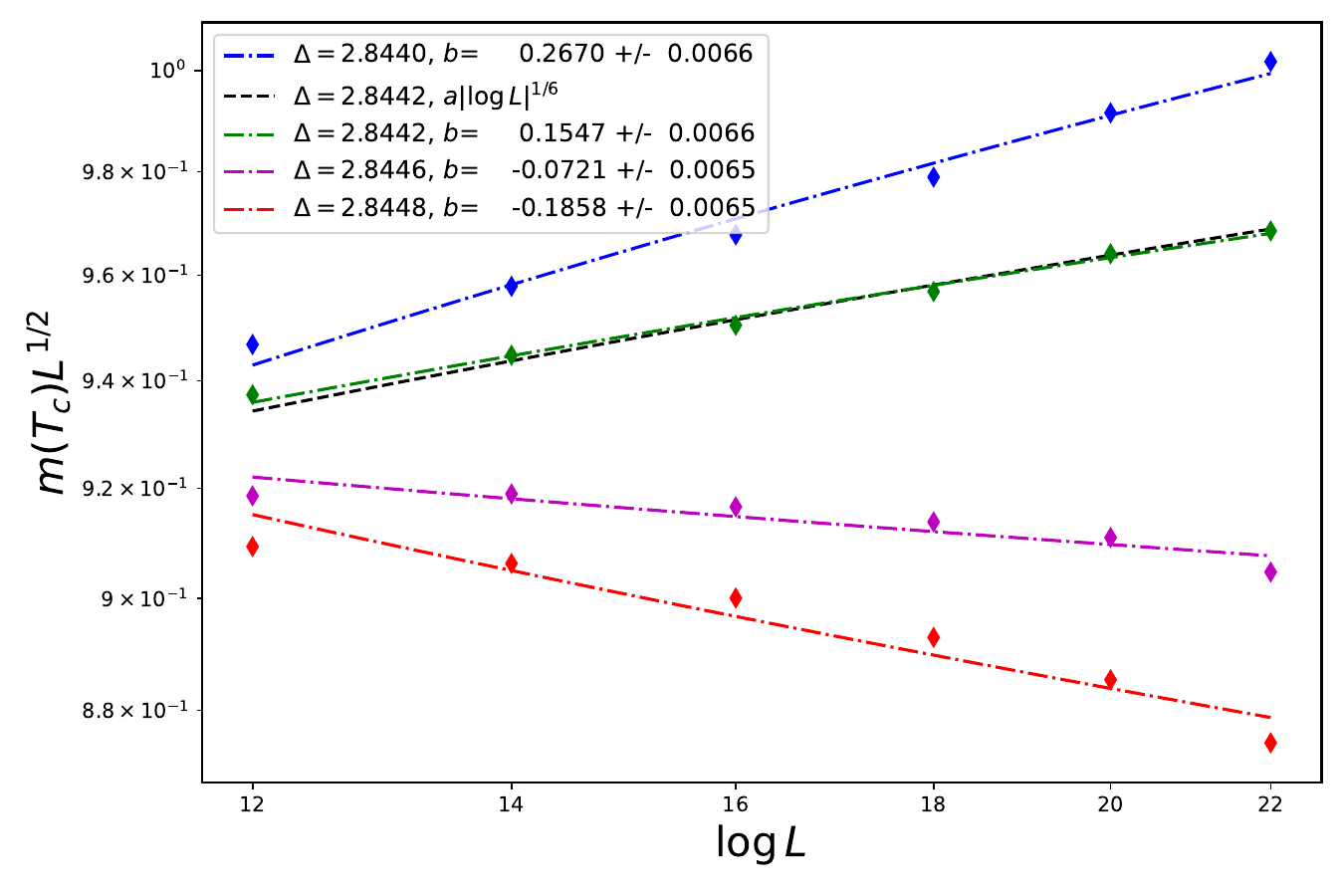} 
\caption{FSS of the magnetization for the tricritical Ising model (Blume-Capel model at its tricritical temperature) in $3d$ at $T_t=1.4182$ and various values of the crystal field parameter $\Delta$ for sizes ranging from $L=12$ to $22$.  The best fit is for $\Delta=2.8442$ ($\chi^2/{\rm dof}= 30.40/4 = 7.6$
at $\Delta=2.8440$,
 $\chi2/{\rm dof}= 5.39/4 = 1.35$ at $\Delta=2.8442$,
$\chi2/{\rm dof}= 28.36/4 = 7.09$ at $\Delta=2.8446$ and
$\chi2/{\rm dof}= 76.74/4 = 19.18$ at $\Delta=2.8448$).
\label{Fig-M-Tc-Delta}}
\end{figure}

As we said, this form can be checked with good accuracy at the price of relatively light Monte Carlo simulations. The coordinates of the tricritical point of the Blume-Capel model in $3d$ are found in the literature~\cite{Deserno}, $T_t\simeq1.4182$, $\Delta_t\simeq 2.84479(30)$, but the value of $\Delta_t$ does not seem to be as accurate as that of the temperature and for example Zierenberg et al~\cite{Zierenberg} report instead $\Delta_t\simeq 2.8446(3)$.
Let us first analyse this problem ourselves. In Figure~\ref{Fig-M-Tc-Delta}, we report the FSS of the magnetization at 
$T_t\simeq1.4182$ for several values of $\Delta$ ranging from 2.8440 to 2.8448. The magnetization is expected to follow the FSS behaviour
\be
m(0,0,L^{-1})\sim L^{-d+y_h} (\ln L)^{\hat y_h}
\ee
with $d-y_h=\frac 12$ and $\hat y_h=\frac 16$. The data points
are fitted as $m(0,0,L^{-1}) L^{\frac 12}=a (\ln L)^b$ with $a,b$  free parameters. The closest to the expected result (the black dashed line) is at $\Delta=2.8442$ where we get $b=0.155\pm 0.007$. We will thus consider this value of $\Delta$ as our estimate for the coordinate of the tricritical point $\Delta_t$.

 \begin{figure}[h!] \centering
\includegraphics[width=0.8\linewidth]{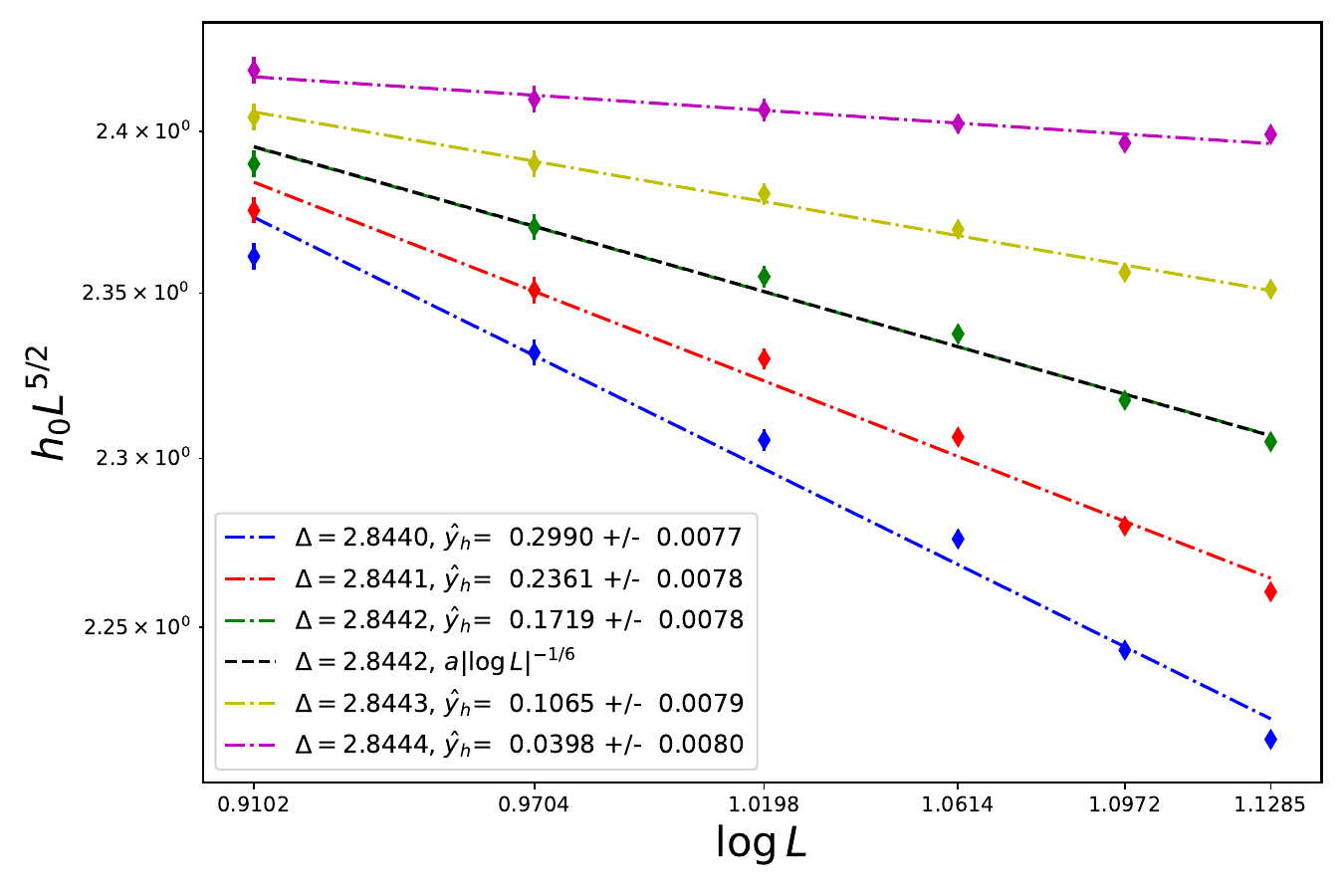} 
\caption{FSS of the Lee-Yang edge for the tricritical Ising model (Blume-Capel model at its tricritical temperature) in $3d$ at $T_t=1.4182$ and various values of $\Delta$ for sizes ranging from  $L=12$ to $22$.\label{Fig-h-vs-L-TriIM}}
\end{figure}

The analysis of the Lee-Yang edge is presented in figure~\ref{Fig-h-vs-L-TriIM} with a larger choice of values of $\Delta$ and, again, the best fit is at $\Delta_t=2.8442$ where the estimate of $\hat y_h$ is now slightly larger at $0.172\pm 0.008$.

\begin{figure}[ht] \centering
\includegraphics[width=0.8\linewidth]{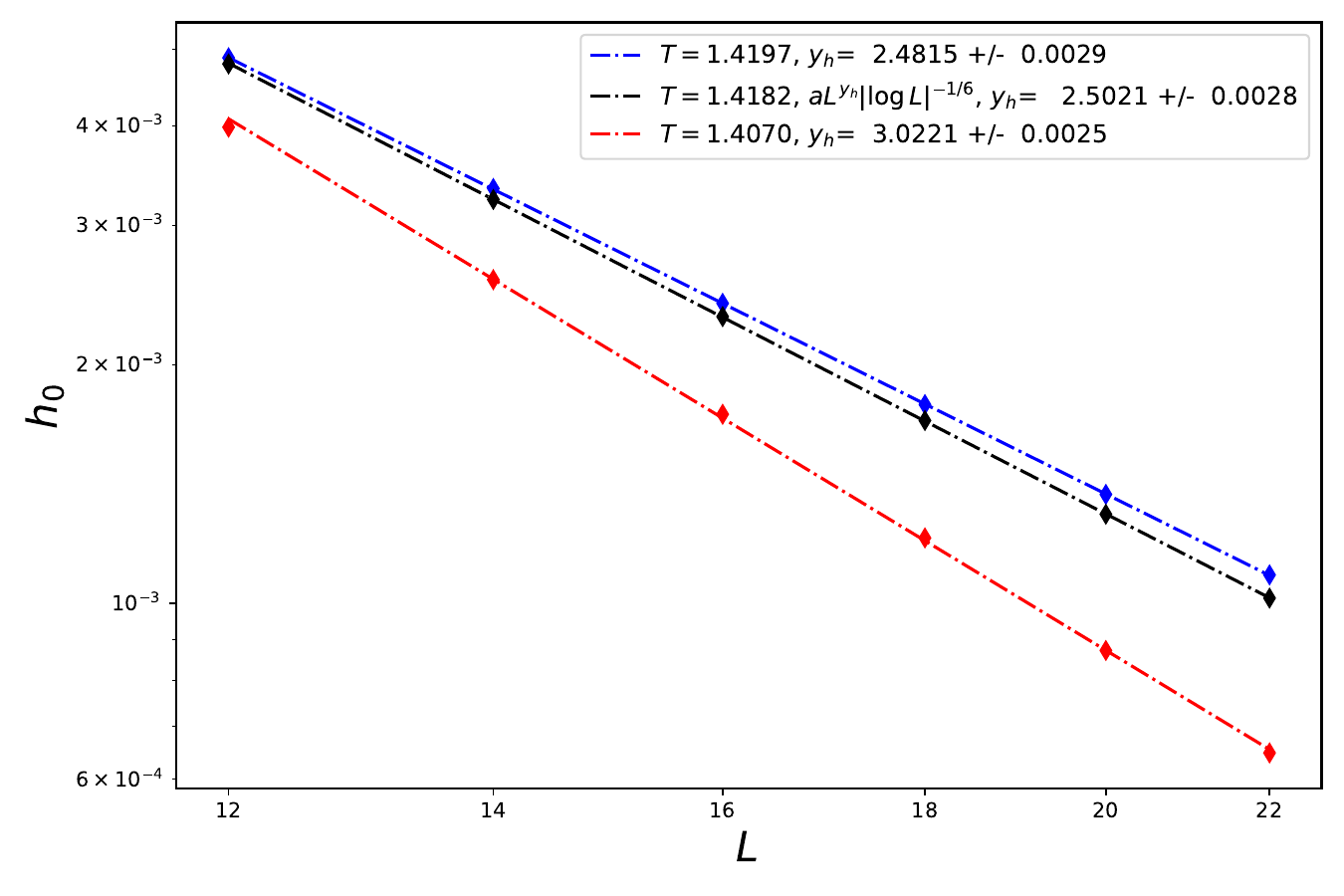} 
\caption{FSS of the Lee-Yang edge in the vicinity of the tricritical point of the Blume-Capel model in $3d$, at $T_t=1.4070$, 1.4182 and 1.4197 and $\Delta=2.8442$ to show the first-order, the tricritical, and the ordinary second-order regimes from the values of the corresponding RG dimensions $y_h$.\label{Fig-h-vs-L-crossover}}
\end{figure}

 The reader could still question the sensitivity of the value of $\hat y_h$ with the choice of tricritical temperature $T_t$. Indeed, when one looks at the FSS of the tricritical magnetization, for example, the effective exponent of the log term is either positive and close to the expected value or can differ from the expectation and even be negative, depending on the values of the crystal field $\Delta$ (see Figure~\ref{Fig-M-Tc-Delta}). It makes sense to ask whether the role of $T_t$ may also have a significative influence.
We believe that the results presented in this work are reliable and to support the consistency of the numerical data, we show in figure~\ref{Fig-h-vs-L-crossover} that slight variations of $T$ change the regime from the pure 3d Ising model at $T = 1.4197$, for which $y_h={2.4815(15)}$ \cite{PV} is expected, to first-order at $T=1.4070$ where an effective FSS $y_h=d$ is expected~\cite{Privman}. $T=1.4182$ safely recovers $y_h^{\rm Tri}=2.5$ to a very good accuracy and confirms the tricritical value of  $T_t\simeq 1.4182$.  Note that the transition line in the phase diagram in the vicinity of the tricritical point is almost at a fixed value of $\Delta$, this is why the three regimes are found at the same crystal field value of $\Delta=2.8442$.

\section{Conclusions}

The numerical results obtained for the tricritical Ising model universality class in $3d$ confirm the prediction that $\hat y_h=\frac 16$, hence the prediction $\hat\eta=\frac 13$, while the scaling law of Kenna and co-workers would have given $-\frac 13$ instead.
  
 In Ref.~\cite{RK3}, Ralph Kenna concluded his review with a table collecting the sets of critical exponents and hatted critical exponents for various models and predicting those which were still unknown from the use of the newly discovered scaling laws, and in particular Eq.~(\ref{eq-RKhat-eta}) that we scrutinize and propose to replace by Eq.~(\ref{eq-RKhat-etamod}) or any of the equivalent forms that we have derived.
 
 \begin{description}
\item In the list, the $O(n)$ model with long-range interactions~\cite{LR1,LR2} was predicted to have $\hat\eta=0$. We propose instead $\hat \eta=\frac 12$, following from $\hat y_t=(4-n)/[2(n+8)]$, $\hat y_h=\frac 14$.

\item The Lee-Yang edge in $6d$~\cite{Ruiz-Lorenzo}  was predictedto have $\hat\eta=\frac 19$. We rather have $\hat y_t=-\frac 29$ and $\hat y_h=\frac 29$, hence $\hat \eta=\frac 49$.

\item For lattice animals in $8d$~\cite{Ruiz-Lorenzo}, Kenna predicted $\hat\eta=\frac 19$ and we have $\hat y_t=\hat y_h=\frac 29$ and $\hat\eta=\frac 49$.

\item The case of scale-free networks~\cite{Net0,Net1,Net2} is particular in the sense that Ralph Kenna did not make any prediction for $\hat\eta$, because some exponents were missing. From those which are known, we can deduce that $\hat y_t=-\frac 12$ and $\hat y_h=-\frac 14$ and we deduce thus $\hat\eta=-\frac 12$ which is a new prediction.

\item Eventually, we believe that the $n$-colour Ashkin-Teller model in $2d$ is still under question since the exponents collected by Shalaev and Jug~\cite{Jug} do not satisfy the ``standard'' scaling laws, e.g. the values reported do not obey $\hat\alpha+\hat\gamma=2\hat\beta$. 
 
 \end{description} 
  
To finish this paper, we would like to say that the scaling laws discovered by Ralph Kenna and his co-workers are invaluable because they make it possible to establish (or falsify) the consistency of the results obtained for various models. The case of the $n$-colour Ashkin-Teller model in $2d$ is such an example where it seems that there are still some inconsistencies to solve. Although we happened to 
contradict one of these scaling laws, we admire the piece of work done in Refs.~\cite{RK3, RK1, RK2} where we recognize Ralph's footprint.


\vspace{1cm}
\noindent
{\bf{Acknowledgements:}}
We are thankful to Wolfhard Janke for very interesting discussions.  
We thank the Coll\`ege Doctoral 02-07 Statistical Physics of Complex Systems (University of Leipzig-Universit\'e de Lorraine) and the L$^4$ collaboration Leipzig - Lviv - Lorraine - Coventry. LM thanks Coventry University and the Coll\`ege Doctoral for financial support.


\end{document}